\documentstyle[12pt]{article}
\input{psfig}
\def\pom{\bf I\! P}
\begin{document}

\title{Pomeron Flux Factor and Diffractive W and Jet Rates
\footnote{Talk presented at the Workshop on Diffractive Physics LISHEP 98, Rio de Janeiro, February 1998.}}
\author{R. J. M. Covolan and M. S. Soares\\
{\small Instituto de F\'{\i}sica {\em Gleb Wataghin}} \\ {\small Universidade
Estadual de Campinas, Unicamp} \\ {\small 13083-970 \  Campinas \  SP \
Brazil \bigskip}}

\date{}

\maketitle

\begin{abstract}
Experimental rates of W and dijets diffractively produced at the Tevatron
Collider have recently become available. We use parametrizations of the pomeron structure
function obtained from  HERA data by two different schemes to compare
theoretical expectations with the measured rates. 
\end{abstract}

\section{Introduction}
The Ingelman-Schlein (IS) model \cite{ingsc}, the first approach proposing the
idea of hard diffraction, predicted that dijets could be produced in $\bar{p}p$ diffractive interactions. This kind of reaction was supposed to occur as a two-step process in which:

1) a pomeron is emitted from the quasi-elastic vertex;

2) partons of the pomeron interact with partons of the incoming 
proton producing jets.

\noindent One notes that the first step refers to a soft process while the second one is typically hard. In the expression proposed to calculate the dijet diffractive cross-section,  the interplay between the soft and hard parts is simply conceived as a product. One assumes that factorization 
applies to these two steps so that

\begin{equation}
\frac{d^{2}\sigma_{jj}}{dtdx_{\pom}}=\frac{d^{2}\sigma_{sd}}{dtdx_{\pom}}
\ \frac{\sigma_{p \pom \rightarrow jj}}{\sigma_{p \pom \rightarrow X}},
\label{sigjj} 
\end{equation}
where $d^{2}\sigma_{sd}/{dtdx_{\pom}}$ is the cross section for single diffraction with $x_{\pom}=M^2_X/s$. 

The soft term in Eq.(\ref{sigjj}),
$(d^{2}\sigma_{sd}/dtdx_{\pom})(1/\sigma_{p \pom \rightarrow X}),$
has become known as the {\it pomeron flux factor} and is usually obtained from the Triple Pomeron Model \cite{pdbcollins} while $\sigma_{p \pom \rightarrow jj}$, the cross section pomeron-proton leading to dijets, is calculated from the parton model and QCD. In order to perform these calculations one has to know the pomeron structure function. This is the subject of Section 2.

The idea of hard diffractive production proposed by the IS model gave rise to a new branch of hadron physics, inspiring a lot of phenomenological studies as well as motivating projects of new experiments. On the phenomenological side,  this concept was extended to processes like diffractive production of heavy flavours, $W$, $Z$, Drell-Yan pairs  (see, for instance, refs.\cite{fritz,DL}). In \cite{DL} appears the 
suggestion of the flux factor as a ``distribution of pomerons in the proton". A particular form of the standard flux is given there by 
\begin{equation}
f(x_{\pom},t)=\frac{9b^{2}}{4\pi^{2}}[F_{1}(t)]^{2}x_{\pom}^{1-2\alpha(t)},
\label{dl} 
\end{equation}
which is usually referred to as Donnachie-Landshoff flux factor.

Most of these processes were incorporated at the events generator POMPYT 
created  by 
Bruni and Ingelman \cite{bruni}. A more recent analysis of diffractive dijets and W production can be found in \cite{collins,jim}. 

As for experimental results, the UA8 Collaboration has recorded the first  observations of diffractive jets \cite{ua8}. This success has inspired other  experimental efforts in this direction. However, subsequent analysis revealed a disagreement between data and theoretical predictions which was referred to as  ``discrepancy factor". 

Goulianos has suggested \cite{dino} that this discrepancy factor observed in hard  diffraction has to do with the well known unitarity violation that occurs
in  soft diffractive dissociation and that it is caused by the flux factor given by the Triple Pomeron Model. In order to overcome this difficulty, he proposed a procedure \cite{dino} which consists of the flux factor ``renormalization", that is, the {\it renormalized} flux is defined as
\begin{equation}
f_{N}(x_{\pom},t)=\frac{f(x_{\pom},t) }{N(x_{\pom_{min}})},
\label{ren}
\end{equation}
where
\begin{equation}
N(x_{\pom_{min}})=\int_{x_{\pom_{min}}}^{x_{\pom_{max}}}dx_{\pom} 
\int^{t=0}_{t=-\infty}f(x_{\pom},t)dt.
\end{equation}

Meanwhile, new data coming from HERA experiment has put the problem of the pomeron structure function in much more precise basis by measuring the {\it diffractive} structure function, {\it i.e.} the proton structure function tagged with rapidity gaps \cite{h1} - \cite{novos}. More recently yet, new diffractive production rates has become available  from experiments performed at the Tevatron by the CDF and D0 collaborations \cite{wcdf} - \cite{d0180}.

This paper consists of a phenomenological analysis in which theoretical predictions  of dijets and W diffractively produced  are presented and compared to the experimental rates. These predictions take for the pomeron structure function results of an analysis performed previously \cite{nois} by using  HERA data.
In such an analysis both possibilities of flux factor, standard and renormalized, are considered.

\section{Pomeron Structure Function from HERA }

The measurements of {\it diffractive} deep inelastic scattering performed by the H1 and the ZEUS collaborations \cite{h1,zeus} at HERA experiment are given in terms of the diffractive cross section
\begin{equation}
\frac{d^{4}\sigma_{ep \rightarrow epX}}{dxdQ^{2}dx_{ \pom}dt}=\frac{4\pi
\alpha^{2}}{xQ^{4}}[1-y+\frac{y^{2}}{2[1+R^{D}(x,Q^{2},x_{ \pom},t)]}] F^{D(4)}_{2}(x,Q^{2},x_{\pom},t),
\end{equation}
where $F^{D(4)}_{2}(x,Q^{2},x_{\pom},t)$ is the {\it diffractive} structure function (details on the notation and kinematics can be found in \cite{nois}).
In these measurements $R^{D}$ was neglected and $t$ was not measured, so that the obtained data were given in terms of
\begin{equation}
F^{D(3)}_{2}(Q^{2},x_{ \pom}, \beta)= \int{F^{D(4)}_{2}(Q^{2},x_{\pom},
 \beta,t)\ dt}.
\end{equation}
The diffractive pattern exhibited by the $F^{D(3)}_{2}$ data \cite{h1,zeus} strongly suggested that the following factorization would apply,
\begin{equation}
F_{2}^{D(3)}=g(x_{ \pom})\ F_{2}^{ \pom}(\beta, Q^{2}).
\label{Fdiff}
\end{equation}
This property is not revealed by data obtained more recently in a extended kinematical region by the H1 Collab. \cite{novos}, but in such a case the violation of factorization basically takes place in the region not covered by the previous measurements and can be attributed to the existence of other contributions besides the pomeron.

Based on the IS model, one can interpret the quantities given in the above equation as $g(x_{ \pom})$, the integrated-over-$t$ flux factor, and $F_{2}^{ \pom}$,  the pomeron structure function.

Our procedure to extract $F_{2}^{\pom}$ from HERA data is basically the following \cite{nois}:

\begin{itemize}

\item{We assume that the factorized expression (\ref{Fdiff}) (and, consequently, pomeron dominance) applies to the kinematical range covered by data given in \cite{h1,zeus};}

\item{For the integrated flux factor, that is for  $g(x_{ \pom}) = \int f(x_{ \pom}, t)\ dt$, both forms, the standard (\ref{dl}) and the  renormalized one (\ref{ren}), are considered;}

\item{The pomeron structure function is given by $ F_{2}^{\pom}(\beta,Q^{2})=\sum_{i} e_{i}^{2}\ \beta q(\beta, Q^{2}) = 2/9\ S(\beta, Q^2)$, where $S(\beta, Q^2) = \sum_{i=u,d,s} \ [q_i(\beta, Q^2) + {\bar q_i} (\beta, Q^2)]$ with $q_{u, {\bar u}} = q_{d, {\bar d}} = q_{s, {\bar s}}$;}

\item{The quark and gluon distribution are evolved in $Q^2$ from a initial scale by the DGLAP equations;}

\item{For the distributions at initial scale $Q^{2}_{0}=4$ GeV$^2$, three possibilities were considered:}

\end{itemize}

\noindent{\bf{1) hard-hard}}:
\begin{eqnarray*}
S(\beta,Q^{2}_{0})&=&a_{1}\ \beta\ (1- \beta)\\
g( \beta,Q^{2}_{0})&=&b_{1}\ \beta\ (1- \beta)
\end{eqnarray*}

\noindent{\bf{2) hard-free:}}
\begin{eqnarray*}
S( \beta,Q^{2}_{0})&=&a_{1}\ \beta\ (1- \beta)\\
g( \beta,Q^{2}_{0})&=&b_{1}\ \beta^{b_{2}}\ (1- \beta)^{b_{3}}
\end{eqnarray*}

\noindent{\bf{3) free-delta:}}
\begin{eqnarray*}
S( \beta,Q^{2}_{0})&=&a_{1}\ \beta^{a_{2}}\ (1- \beta)^{a_{3}}\\
g( \beta,Q^{2}_{0})&=&b_{1}\ \delta(1- \beta).
\end{eqnarray*}

The detailed description of these fits and results can be found in \cite{nois}.
Since for the case of renormalized flux it was difficult to establish the gluon component, a fourth possibility was used in which  the initial distribution of gluons was supposed to be null. Four of these fits were selected from \cite{nois} to perform the calculation of the diffractive rates presented here.
The parameters used in such calculations are shown in Table \ref{um}.

\begin{table}
\begin{center}
\begin{tabular}{ccccc}
\hline \hline \\
& D\&L & D\&L & REN & REN \\
& hard-hard & free-delta & hard-hard & free-zero\\ \hline \\
$a_1$ & 2.55 & 1.51 & 5.02 & 2.80\\
$a_2$ & 1 & 0.51 & 1 & 0.65\\
$a_3$ & 1 & 0.84 & 1 & 0.58\\
$b_1$ & 12.08 & 2.06 & 0.98 & $-$ \\
$b_2$ & 1 & $-$ & 1 & $-$ \\
$b_3$ & 1 & $-$ & 1 & $-$ \\ 
\hline \hline
\end{tabular}
\vspace{0.2cm}
\caption{\sf{Fit parameters for the pomeron structure function. The procedure used to establish these parametrizations can be found in {\protect{\cite{nois}}}.}} 
\label{um}
\end{center}
\end{table}


\section{Diffractive Parton Model}

In this section, we present the expressions we have used to calculate the rates for diffractive production of W and jets. From the parton model, the generic expression for the cross section of these processes is 

\begin{equation}
\label{partonmodel}
d \sigma_{W/jj} = f_A(x_a, Q^2)dx_a\ f_B(x_b,Q^2)dx_b\ (d{\hat{\sigma}}_{ab})_{W/jj}
\end{equation}
where the parton {\it a} of the hadron {\it A} interacts with the parton {\it b} of the hadron {\it B} to give a W or a pair of partons $(c, d)$ in the case of dijets.

\subsection{W production}

With the elementary cross section 
\begin{equation}
{\hat{\sigma}_{ab\rightarrow W}}= \frac{2}{3} \pi {g_W}^2\ \delta(x_a\ x_b\ s-M^2_W)
\end{equation}
in equation (\ref{partonmodel}), the integrated cross section is given by

\begin{equation}
\sigma(AB \rightarrow W^{\pm})= \frac{2}{3} \pi \frac{{g_W}^2}{s} \sum_{a, b}  \int_{x_{a_{min}}}^{1}\frac{dx_a}{x_a}
f_A(x_a)\ f_B(x_b).
\label{www}
\end{equation}
with $x_b={M^2_W}/{x_a\ s}$.
For $W^+$ production, the interacting partons are  $a=u$ and 
$b={\bar d}_{\theta_C}$, and for $W^-$ production, $a=\bar{u}$ and $b=d_{\theta_C}$, with $d_{\theta_C}=d\cos{\theta_C}+s\sin{\theta_C}$ where $\theta_C$ is the Cabbibo angle ($\theta_C \cong 13^o$). The kinematical limit is determined by 
$x_a\ x_b\ s = \hat{s}=M^2_W$, that is $x_{a_{min}} = {M^2_W}/{s}$.

\subsection{Dijets production}

In the case of dijets generated from partons {\it c} and {\it d}, their 
transversal energy is  

\begin{equation}
E_T = |p_c|\sin{\theta_c} = |p_d|\sin{\theta_d}.
\end{equation}
By using the definition of rapidity, 

\begin{eqnarray*}
y=\frac{1}{2} \ln{\frac{E+E_L}{E-E_L}}
\end{eqnarray*}
one can get
\begin{equation}
e^{-y}=\frac{E_T}{|p_c|(1+\cos{\theta_c})}
\ \ \ \ {\rm and} \ \ \ \ 
e^{-y'}=\frac{E_T}{|p_d|(1+\cos{\theta_d})}.
\end{equation}
Defining the Mandelstam variables for the parton system as
\begin{equation}
\hat{s}=x_a x_b s 
\end{equation}
and
\begin{eqnarray*}
\hat{t}=-2p_a.p_c &=& -x_a \ \sqrt{s}\ E_T\ e^{-y}
= -x_b\ \sqrt{s}\ E_T\  e^{y'},
\end{eqnarray*}
one can write the Bjorken variables $x_a$ and $x_b$ as

\begin{equation}
x_a=\frac{E_T}{\sqrt{s}}(e^{y}\ +\ e^{y'})
\ \ \ \ 
{\rm and} \ \ \ \ 
x_b=\frac{E_T}{\sqrt{s}}(e^{-y}\ +\ e^{-y'}).
\end{equation}
Now, making use of the transformation 
$dx_a\ dx_b\ d\hat{t} \rightarrow 2E_T\ dE_T\ x_a\ x_b\ dy'\ dy$ 
in Eq. (\ref{partonmodel}), one obtains

\begin{eqnarray}
\frac{d\sigma}{dy}=\sum_{a, b}\int_{E_{T_{min}}}^{E_{T_{max}}} dE_{T}^2 \int_{y'_{min}}^{y'_{max}} dy' x_a f_A(x_a, Q^2) x_b f_B(x_b,Q^2)(\frac{d\hat{\sigma}}{d\hat{t}})_{jj}.
\label{dijet}
\end{eqnarray}

\noindent In this case, the kinematical limits are

\begin{eqnarray*}
\ln{\frac{E_T}{\sqrt{s}-E_T\ e^{-y}}} \leq &y'& \leq \ln{\frac{\sqrt{s}-E_T\ e^{-y}}{E_T}}, \\ \\
E_{T_{min}}=experimental\ &cut& \ \ \ \ {\rm and} \ \ \ \
E_{T_{max}}=\frac{\sqrt{s}}{e^{-y}+e^{y}}.
\end{eqnarray*}

\subsection{Diffractive Dijets and W production}

In order to calculate the diffractive cross sections, we use the 
Pomeron structure function defined as 

\begin{eqnarray}
xf_{\pom}(x, Q^2)=\int dx_{\pom} \int d\beta\ g(x_{\pom})\ \beta
f_{\pom}(\beta, Q^2) \delta(\beta-\frac{x}{x_{\pom}}).
\end{eqnarray}
Introducing this expression in Eq.(\ref{dijet}), we obtain the cross section
for diffractive dijet production,
\begin{eqnarray}
\frac{d\sigma}{dy}=\sum_{a, b}\int_{E_{T_{min}}}^{E_{T_{max}}} dE_{T}^2  \int_{y'_{min}}^{y'_{max}} dy' \int_{x_{\pom min}}^{x_{\pom max}} dx_{\pom}   g(x_{\pom})\ x_a f_p(x_a, Q^2)\ \beta f_{\pom}(\beta,Q^2)\ (\frac{d\hat{\sigma}}{d\hat{t}})_{jj},
\label{ddijet}
\end{eqnarray}
where the scale is given by $Q^2 = E_T^2$.

As for diffractive W production, the expression obtained is

\begin{eqnarray}
\sigma(p\bar{p} \rightarrow W^{\pm})=\sum_{a, b}\frac{2}{3} \pi \frac{{g_W}^2}{s} \int_{x_{\pom min}}^{x_{\pom max}} dx_{\pom} \int_{\beta_{min}}^{1}\frac{d\beta}{x_{\pom} \beta}\ g(x_{\pom}) 
f_{\pom}(\beta,Q^2) f_p(\frac{\tau}{x_{\pom} \beta},Q^2),
\label{dwww}
\end{eqnarray}
where $\tau=M^2_W / s$ , $\beta_{min}=\tau /x_{\pom}$ and $Q^2 = M^2_W$.

In all of these calculations, the parametrizations used for the proton structure function were taken from ref.\cite{gluck}.

\section{Results and discussion}

The experimental rate for diffractive production of W is \cite{wcdf} $R_W\ =\ (1.15\pm 0.55)\ \%$. Table \ref{dataprod} summarizes the experimental data from  \cite{rapgap,roman,d0180}  referring to the diffractive production rates of dijets as well as the kinematical cuts used to obtain these
data.

\begin{table}
\vspace{-1.5cm}
\begin{tabular}{|c|c|c|c|c|}
\hline \hline 
& & & & \\
CUTS &  CDF (Rap-Gap) & CDF (Roman Pots) & D0 1800 & D0 630 \\ 
 & & & & \\ \hline
 & & & & \\
rapidity & $-3.5 \leq y \leq -1.8$ & $-3.5 \leq y \leq -1.8$ ~ & 
 $-4.1 \leq y \leq -1.6$ ~ & $-4.1 \leq y \leq -1.6$ ~ \\ 
 & & & & \\ \hline 
 & & & & \\
$x_{\pom}$ &  $  x_{\pom} \leq 0.1 $ & $ 0.05 \leq x_{\pom} \leq 0.1$ & $ x_{\pom} \leq 0.1 $ & $  x_{\pom} \leq 0.1 $ \\
 & & & & \\ \hline
 & & & & \\
$E_{T}(min)$ & $20\ GeV$ & $10\ GeV$ &  $12\ GeV$  & $12\ GeV$ \\ 
 & & & & \\ \hline 
 & & & & \\
RATES & $0.75\pm 0.10$ (2j+3j) ~ & $0.109\pm 0.016$ &  & \\
 & & & $0.67\pm 0.05$ & 1-2 \\
(\%) & 1.53* (2j) & (2j) & & \\
 & & & & \\
\hline \hline
\end{tabular}
\vspace{0.3cm}
\caption{\sf{Experimental data of diffractive production of dijets and kinematical cuts. In the first column, the rate includes contribution of a third jet. The number given below indicated with an asterisk is the rate corrected to dijets only.
\label{dataprod}}}
\end{table}

In Figs.1-4, we present the rapidity distributions of jet cross section obtained with different parametrizations for the pomeron structure function
and for both flux factors.

The experimental data are shown again in Table \ref{resultrates} in comparison with the rates obtained from our theoretical calculations. The results obtained with the standard (Donnachie-Landshoff) flux are indicated by D \& L, while the columns indicated as REN give the results obtained with the renormalized (Goulianos) flux. By looking at these results, we can note the following:

\begin{itemize}

\item{The rates obtained with standard flux are much larger than the experimental values, being that these discrepancies are  more pronounced for hard gluon distributions;}

\item{Generally speaking, the rates obtained with the renormalized flux are very close to the experimental data;}

\item{The experimental rate for the case of dijets-CDF obtained with rapidity gaps increases when one excludes the contamination with the third jets (see Table \ref{dataprod} , second column); thus we see that the renormalized flux generally underestimates the rates except for the case of jets-CDF obtained with roman pots, in which the contrary happens;}

\item{A lack of W's is noticed in the renormalized case in spite of the fact that the pomeron structure function for this case implies that 
the quark component is pratically the double of the gluon component \cite{nois}.} 

\end{itemize}

\section{Concluding remarks}

The results of W and dijet production rates presented in this paper show that, in order to make the theoretical predictions obtained with the pomeron structure function extracted from HERA data compatible with experimental data of such rates, a renormalization procedure (or something alike) is indispensable. Of course, this conclusion is conditioned by the presumptions that underlie the approach used here, that is the Ingelman-Schlein model.

\section*{Acknowledgement}

We would like to thank the Brazilian governmental agency FAPESP for the financial support.

\begin{table}
\begin{tabular}{|c|c||c|c|c|c|} 
\hline \hline 
 &  &  &  &  & \\
& & D \& L  &  D \& L & REN & REN \\
EXPERIMENT & RATES &  &  &  &  \\
 & & hard-hard ~ & free-delta ~ & hard-hard ~ & free-zero ~ \\
 &  & &  &  &  \\ \hline
& & & & & \\ 
{\it Jets} - CDF (Rap-Gap) & $0.75\pm 0.10$  & 15.3  & 6.33 & 0.62 & 0.52  \\   & (2j+3j) & & & & \\  & & & & & \\ \hline
& & & & & \\ 
{\it Jets} - CDF (Roman Pots)  & $0.109\pm 0.016$ ~ & 3.85 & 1.13 & 0.15  &  0.16 \\  & & & & & \\ \hline  & & & & & \\ 
{\it Jets} - D0 $630\  GeV$ & 1-2 & 15.4 & 6.41 & 0.87 & 0.71  \\ 
 & & & & & \\ \hline  & & & & & \\ 
{\it Jets} - D0 $ 1800\  GeV$  & $0.67\pm 0.05$  & 16.6 & 6.14 & 0.65 & 0.57  \\  & & & & & \\ \hline    & & & & & \\ 
{\it W's} - CDF (Rap-Gap) & $1.15\pm 0.55$ & 3.12 & 3.54 & 0.53 & 0.58   \\ 
  & & & & & \\
\hline \hline
\end{tabular}
\vspace{0.2cm}
\caption{\sf{Production rates - all values are given in percentages.}}
\label{resultrates}
\end{table}

\newpage

\twocolumn

\begin{figure}
\vspace{-2cm}
\centerline{\psfig{figure=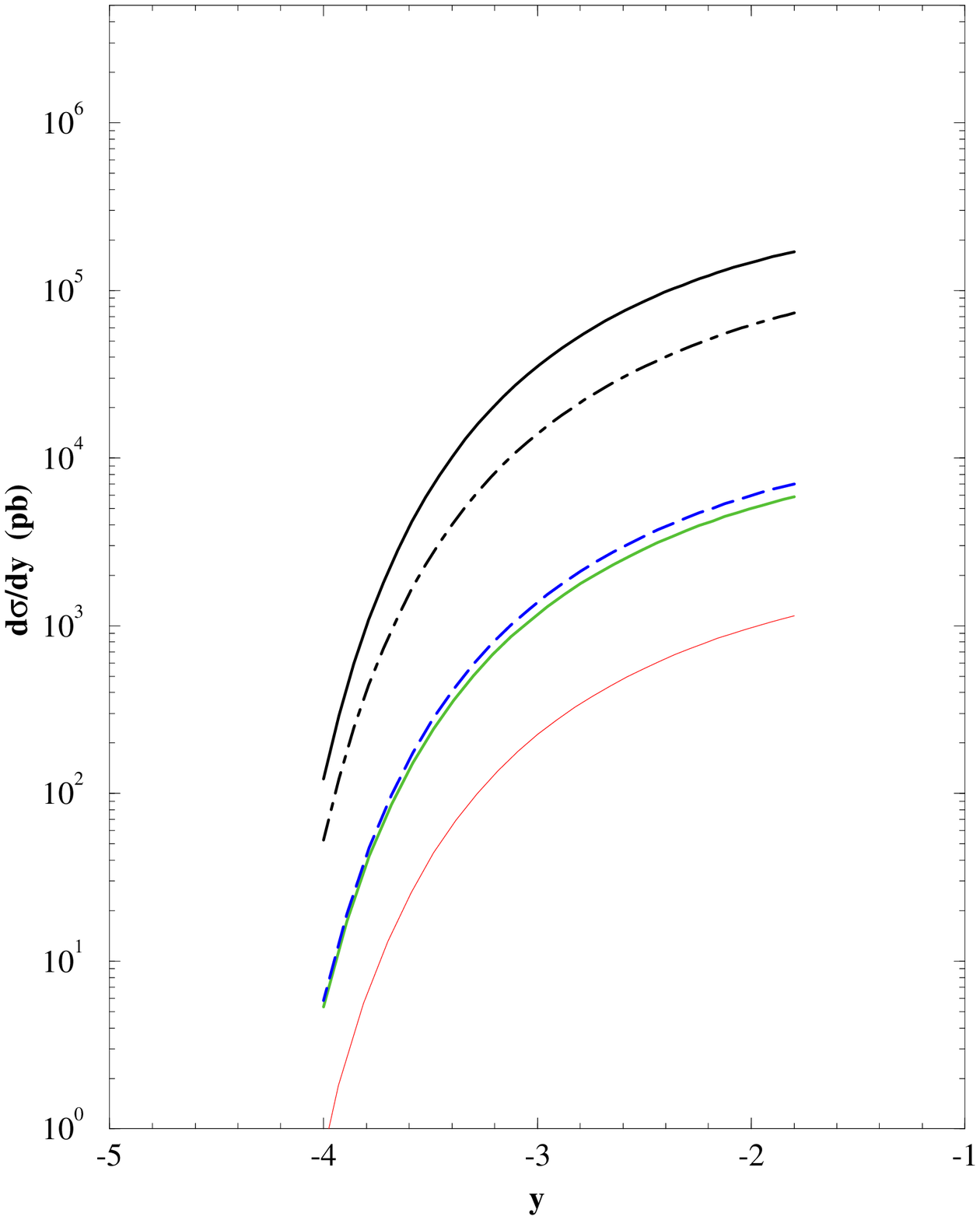,width=7.6cm,height=8cm}}
\caption{Rapidity distribution of dijet production. Kinematical cuts corresponding to the CDF experiment \protect{\cite{rapgap}}.\label{rapgap_fig}
}
\end{figure}

\begin{figure}[htbp]
\centerline{\psfig{figure=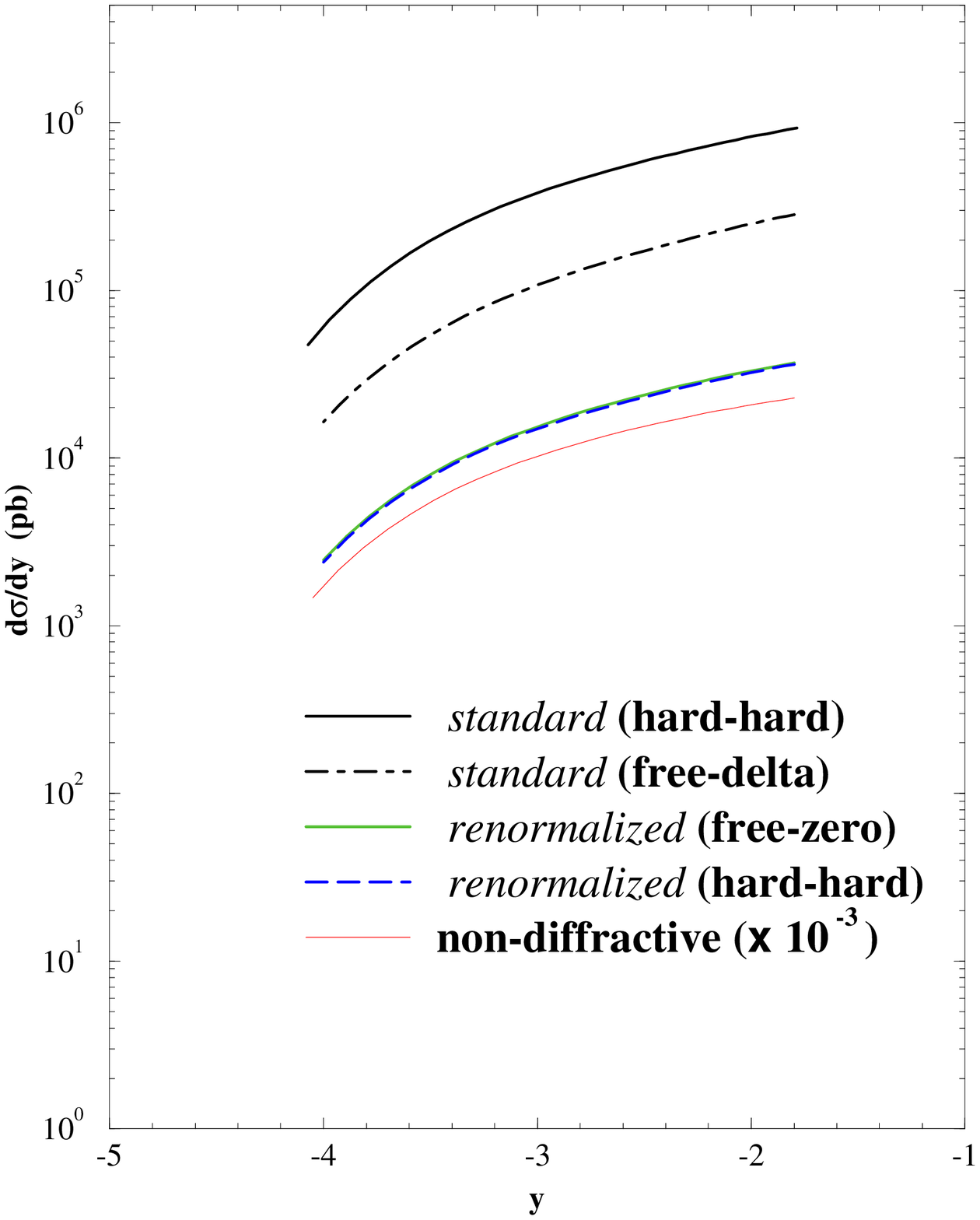,width=7.6cm,height=8cm}}
\caption{Rapidity distribution of dijet production. Kinematical cuts corresponding to the CDF experiment \protect{\cite{roman}}.}
\end{figure}

\newpage

\begin{figure}[htbp]
\vspace{-2cm}
\centerline{\psfig{figure=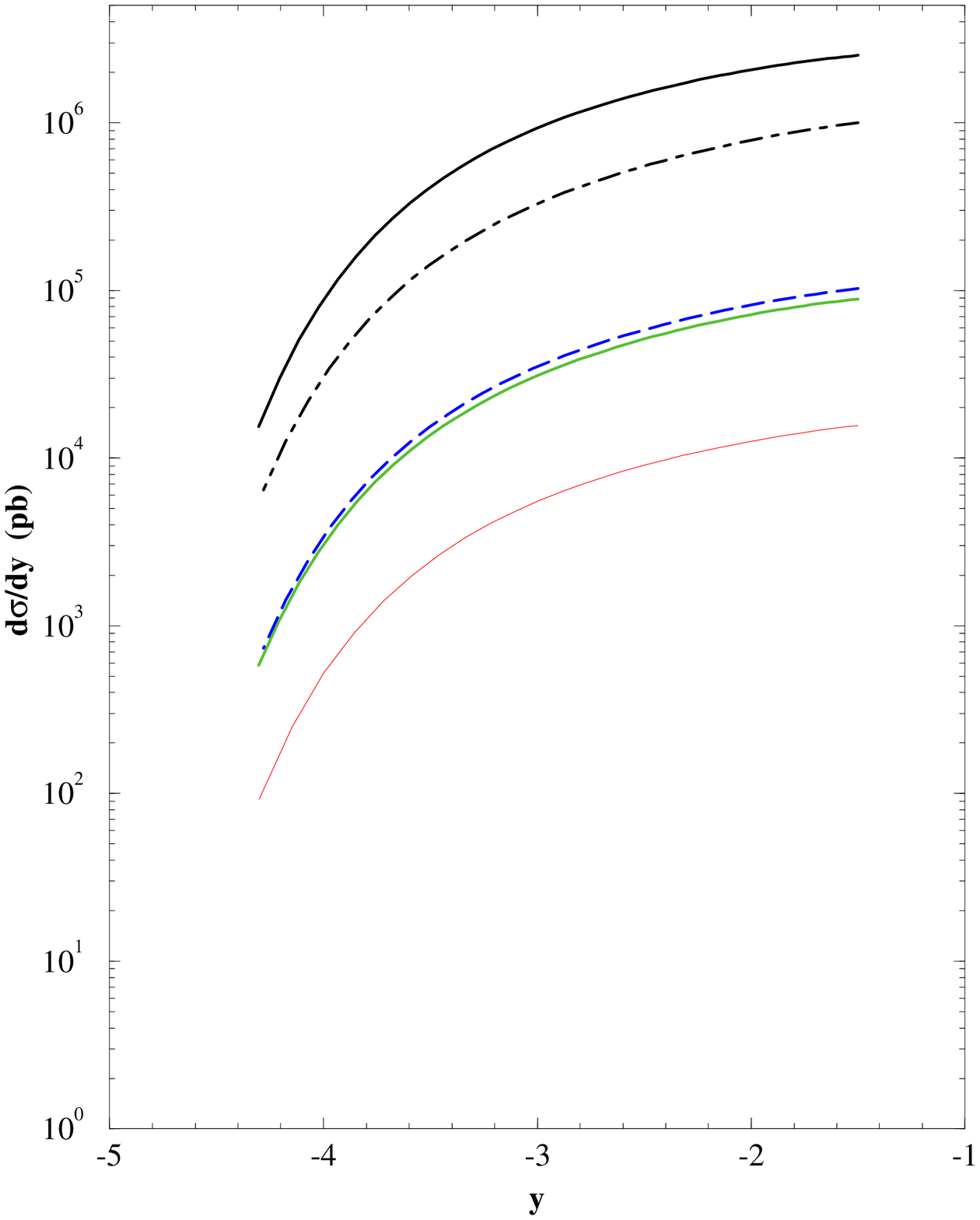,width=7.6cm,height=8.cm}}
\caption{Rapidity distribution of dijet production for 1800 GeV. Kinematical cuts corresponding to the D0 experiment \protect{\cite{d0180}}.}
\end{figure}

\begin{figure}[htbp]
\centerline{\psfig{figure=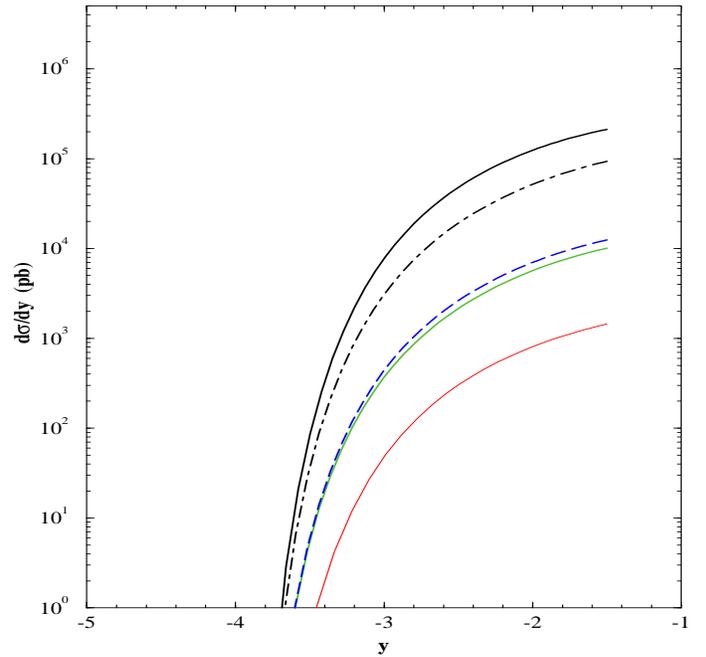,width=7.6cm,height=8.cm}}
\caption{Rapidity distribution of dijet production for 630 GeV. Kinematical cuts corresponding to the D0 experiment \protect{\cite{d0180}}.}
\label{d0630_fig}
\end{figure}


\begin{thebibliography}{99}

\bibitem{ingsc} G.Ingelman, P.E.Schlein, Phys. Lett. {\bf B 152}, 256 (1985).

\bibitem{pdbcollins} P. D. B. Collins, {\it An Introduction to Regge Theory 
and High Energy Physics}, (Cambridge University Press, Cambridge, 
England, 1977).

\bibitem{fritz} H. Fritzsch, K.-H. Streng, Phys. Lett. {\bf B 164}, 391 (1985).

\bibitem{DL} A. Donnachie, P.V.Landshoff, Nucl. Phys. {\bf B 303}, 634 (1988).

\bibitem{bruni} P.Bruni, G.Ingelman, Phys. Lett. {\bf B 311}, 317 (1993); preprint DESY 93-187 (1993).

\bibitem{collins} L.Alvero, J.C.Collins, J. Terron, J.Whitmore, preprint {\bf PSU/TH/177 }, hep-ph/9805268. 

\bibitem{jim} J. Whitmore, Proceedings of the  Workshop on Diffractive Physics LISHEP 98, Rio de Janeiro, February 1998.

\bibitem{ua8} R. Bonino {\it et al.} (UA8 Collaboration), Phys. Lett. {\bf B 211}, 239 (1988). 

\bibitem{dino} K. Goulianos, Phys. Lett. {\bf B 358}, 379 (1995). 

\bibitem{h1} T. Ahmed {\it et al.} (H1 Collaboration), Phys. Lett. {\bf B 348}, 681 (1995). 

\bibitem{zeus}  M. Derrick {\it et al.} (ZEUS Collaboration), Zeit. Phys. 
{\bf C 68}, 569 (1995). 

\bibitem{novos} C. Adloff {\it et al.} (H1 Collaboration), Zeit. Phys. {\bf C 76}, 613 (1997). 

\bibitem{wcdf} F. Abe {\it et al.} (CDF Collaboration), Phys.Rev.Lett. {\bf 78}, 2698 (1997). 

\bibitem{rapgap} F. Abe {\it et al.} (CDF Collaboration), Phys. Rev. Lett. {\bf 79}, 2636 (1997). 

\bibitem{roman} M. G. Albrow, Proceedings of the VII Blois Workshop (to be published). 


\bibitem{d0180} S. Abachi {\it et al.} (D0 Collaboration), paper presented at 
the 28th Int. Conf. on High Energy Physics (ICHEP 96), Warsaw, Poland 
(July 1996).

\bibitem{nois} R.J.M.Covolan, M.S.Soares, Phys. Rev. {\bf D 57}, 180 (1998). 

\bibitem{gluck} M.Gluck, E.Reya, A.Vogt, Zeit. Phys. {\bf C 67}, 433 (1995).

\end{thebibliography}
\end{document}